\documentclass[10pt,english,floatfix,superscriptaddress,aps,prd,preprint,showkeys,nofootinbib]{revtex4}
\usepackage{amsmath}
\usepackage{amssymb}
\usepackage{amsbsy}
\usepackage{amsfonts}
\usepackage{amsopn}
\usepackage{amstext}
\usepackage{graphicx}
\usepackage{amssymb}
\usepackage{amsfonts}
\usepackage{amsmath}
\usepackage{graphicx}
\usepackage[english]{babel}
\usepackage{color}
\usepackage{slashed}
\usepackage{esint}
\usepackage[dvips]{epsfig}
\usepackage[dvips]{graphicx}
\usepackage{float}
\usepackage{units}
\usepackage{textcomp}
\usepackage{placeins}

\usepackage{hyperref}
\usepackage{slashed}

\begin{document}

\title{Gravitational waves effects in a Lorentz--violating scenario}


\author{K. M. Amarilo}
\email{kevin.amarilo@cern.ch}
\affiliation{Dep. de Física Nuclear e Altas Energias, Instituto de Física, Universidade do Estado do Rio de Janeiro, Rua São Francisco Xavier, 524, BR - Rio de Janeiro 20559-900, RJ, Brazil.}


\author{M. B. Ferreira Filho}
\email{mapse.b@cern.ch}
\affiliation{Dep. de Física Nuclear e Altas Energias, Instituto de Física, Universidade do Estado do Rio de Janeiro, Rua São Francisco Xavier, 524, BR - Rio de Janeiro 20559-900, RJ, Brazil.}


\author{A. A. Ara\'{u}jo Filho}
\email{dilto@fisica.ufc.br}
\affiliation{Departamento de Física Teórica and IFIC, Centro Mixto Universidad de Valencia--CSIC. Universidad de Valencia, Burjassot-46100, Valencia, Spain}
\affiliation{Departamento de Física, Universidade Federal da Paraíba, Caixa Postal 5008, 58051--970, João Pessoa, Paraíba,  Brazil.}

\author{J. A. A. S. Reis}
\email{jalfieres@gmail.com}

\affiliation{Universidade Estadual do Sudoeste da Bahia (UESB), Departamento de Ciências Exatas e Naturais, Campus Juvino Oliveira, Itapetinga -- BA, 45700-00,--Brazil}

\date{\today}

\begin{abstract}

This paper focuses on how the production and polarization of gravitational waves are affected by spontaneous Lorentz symmetry breaking, which is driven by a self--interacting vector field. Specifically, we examine the impact of a smooth quadratic potential and a non--minimal coupling, discussing the constraints and causality features of the linearized Einstein equation. To analyze the polarization states of a plane wave, we consider a fixed vacuum expectation value (VEV) of the vector field. Remarkably, we verify that a space--like background vector field modifies the polarization plane and introduces a longitudinal degree of freedom. In order to investigate the Lorentz violation effect on the quadrupole formula, we use the modified Green function. Finally, we show that the space--like component of the background field leads to a third--order time derivative of the quadrupole moment, and the bounds for the Lorentz--breaking coefficients are estimated as well.

\end{abstract}


\maketitle

\section{Introduction}

Understanding the spacetime structure at the Planck scale and developing a corresponding quantum theory of gravity is one of the most challenging open questions in modern physics. At such high--energy scales, it is possible that some of the low--energy symmetries present in the standard model of particles and gravity, such as Lorentz and CPT symmetries, may be violated. This phenomenon has been observed in various theories, including some string theory vacuum \cite{1,1.2}, Loop Quantum Gravity \cite{2,2.1}, non--commutative geometry \citep{3,araujo2023thermodynamics,heidari2023gravitational}, Horava Gravity \citep{4}, and Very Special Relativity \citep{5,furtado2023thermal}.

The Standard Model Extension (SME) \cite{6,6.1} provides a general framework for probing the remnants of Lorentz--violating effects at low--energy regimes. By considering coupling terms between the Standard Model fields and Lorentz--violating coefficients, several studies have been addressed in a different context, such as photons \citep{7,araujo2021thermodynamic,araujo2021higher,araujo2022particles,araujo2021bouncing,araujo2022thermal}, neutrinos \citep{8}, muons \citep{9}, quantum rings \cite{,ww1}, and hydrogen atoms \citep{10}, among others \cite{filho2022thermodynamics,li2017application,cambiaso2012massive,ww2,ww3,ww4}. A comprehensive list of tests for Lorentz and CPT symmetry breaking can be found in Ref. \citep{11}. In addition, in order to incorporate gravity into the Standard Model Extension (SME), we must assume the possibility of spontaneous Lorentz symmetry violation. This is typically achieved by introducing self--interacting tensor fields with a non--null vacuum \cite{bluhm2008spontaneous}.

Gravitational wave detection was one of the most important events for high--energy physics. Two notable events took place in 2015 and were announced on February 11th, 2016 by the VIRGO and LIGO laboratories. The first event, GW150914, was detected on September 14th, 2015 \cite{12}. The second one, GW151226, was detected on December 26th, 2015. Both events were caused by the coalescence of two black holes located at distances of approximately 410 Mpc and 440 Mpc, respectively \cite{13}. The detection of gravitational waves from these events represents a major milestone in the study of black holes and the nature of gravity, opening up new avenues for research and providing important insights into the workings of the universe at the most fundamental level \cite{aa2023analysis,filho2024implications}.

One relatively simple field that breaks the Lorentz symmetry is the bumblebee field $B_\mu$. The Lorentz symmetry is spontaneously broken by the dynamics of $B_\mu$, which acquires a non--zero vacuum expectation value (VEV) \cite{15, 16,maluf2019antisymmetric}. This model was originally considered in the context of string theory, where LSB is triggered by the potential $V(B^\mu) = \lambda(B^\mu B_\mu \mp b^2)^2/2$.  The Einstein–-aether gravity, as outlined in \cite{sagi2010propagation}, represents a general tensor--vector theory at the two--derivative levels. This implies that the considered Lagrangian encompasses generic terms, manifesting quadratically in its derivatives. In the context of bumblebee gravity, the introduction of a vector background denoted as $b^\mu$, with exclusively a nonzero temporal component, engenders five independent propagating degrees of freedom. Notably, a parallel occurrence of this characteristic is observed in the Einstein--aether gravity theory, as discussed in \cite{liang2022polarizations}.

This work focuses on analyzing the modifications in the polarization and production of gravitational waves due to Lorentz symmetry breaking (LSB) in the weak field regime. In particular, we investigate the effects of the spontaneous breaking of Lorentz symmetry, triggered by the bumblebee vector field. Then, we solve the modified wave equation for the perturbation field and compare the polarization tensor of the modified gravitational wave solution with the conventional case. Next we introduce a current $J_{\mu\nu}$ to the model and derive a Green function for the timelike $b^\mu = (b^0, 0)$ and spacelike $b^\mu = (0, {\bf b})$ configurations for the bumblebee VEV. Subsequently, we obtain a modified quadrupole formula for the graviton, which enables us to compare the perturbation for the modified equation with the usual one and to identify the modifications in the theory.


\section{Modifying the Graviton Wave Equation}

The minimal extension of the gravity theory, including the Lorentz--violating terms is given by the following expression \cite{17}
\begin{equation}
    S = S_{EH} + S_{LV} + S_{matter}.
\end{equation}
The first term refers to the usual Einstein--Hilbert action,
\begin{equation}
    S_{EH} = \int \mathrm{d}^4x \; \sqrt{-g} \frac{2}{\kappa^2} (R - 2\Lambda),
\end{equation}where $R$ is the curvature scalar and $\Lambda$ is the cosmological constant, which will not be considered in this analysis. The $S_{LV}$ term consists of the coupling between the bumblebee field and the curvature of spacetime
\begin{equation}\label{eq:SLV}
    S_{LV} =  \int \mathrm{d}^4x \; \sqrt{-g} \frac{2}{\kappa^2} (uR + s^{\mu\nu}R_{\mu\nu} + t^{\alpha\beta\mu\nu} R_{\alpha\beta\mu\nu}),
\end{equation}where $u$, $s^{\mu\nu}$ and $t^{\alpha \beta \mu \nu}$ are dynamical fields with zero mass dimension \cite{nascimento2022vacuum}. And the last term represents the matter--gravity couplings which, in principle, should include all fields of the standard model as well as the possible interactions with the coefficients $u$, $s^{\mu \nu}$ and $t^{\alpha \beta \mu \nu}$.

The dynamics of the bumblebee field $B^\mu$ is dictated by the action \cite{18,hassanabadi2023gravitational}
\begin{equation}
\label{acttion}
    S_B = \int \mathrm{d}^4x \; \sqrt{-g} \left[ -\frac{1}{4} B^{\mu \nu} B_{\mu \nu} + \frac{2 \xi}{\kappa^2} B^\mu B^\nu R_{\mu\nu} - V(B^\mu B_\mu \mp b^2) \right],
\end{equation}where we introduced the field strength $B_{\mu \nu} = \partial_\mu B_\nu - \partial_\nu B_\mu$ in analogy with the electromagnetic field tensor $F_{\mu\nu}$. In fact, the bumblebee models are not only used as toy models to investigate the excitation originating from the LSB mechanism. They also provide an alternative to $U(1)$ gauge theory for photons \cite{19}. In this theory, they appear as Nambu--Goldstone modes due to spontaneous Lorentz violation \cite{20}, rather than fundamental particles.

The gravity--bumblebee coupling can be represented by Eq. (\ref{eq:SLV}) defining
\begin{equation}
    u = \frac{1}{4}\xi B^\alpha B_\alpha, \quad s^{\mu\nu} = \xi \left(B^\mu B^\nu -\frac{1}{4} g^{\mu\nu} B^\alpha B_\alpha \right), \quad t^{\alpha \beta \mu \nu} = 0,
\end{equation}
and the potential $V(X)$ reads
\begin{equation}
V=\frac{\lambda }{2}\left( B^{\mu }B_{\mu }\mp b^{2}\right) ^{2}.
\end{equation}
It is responsible for triggering spontaneously the Lorentz violation and breakdown of the diffeomorphism. Here, $b^2$ is a positive constant that stands for the non--zero vacuum expectation value of this field.
In our study, we aim to investigate how the coupling between gravity and the bumblebee field affects the behavior of graviton. To do so, we consider the linearized version of the metric $g_{\mu \nu}$ with the Minkowski background and the bumblebee field $B_\mu$, which is split into the vacuum expectation values $b_\mu$ and the quantum fluctuations $\Tilde{B}^\mu$
\begin{eqnarray}
g_{\mu\nu} & = & \eta_{\mu\nu}+\kappa h_{\mu\nu},  \notag \\
B_{\mu} & = & b_{\mu}+\tilde{B}_{\mu},  \notag \\
B^{\mu} & = & b^{\mu}+\tilde{B}^{\mu}-\kappa b_{\nu}h^{\mu\nu}.
\label{eq:Expansion1}
\end{eqnarray}

By multiplying the linearized equation of motion with $p^\mu p^\nu$, $b^\mu b^\nu$, $p^{(\mu} b^{\nu)}$, the set of constraints are obtained 
\begin{equation}\label{constrains}
    p_\mu p_\nu h^{\mu\nu} = 0, \quad b_\mu b_\nu h^{\mu\nu} = 0, \quad p_{(\mu} b_{\nu)} h^{\mu\nu} = 0, \quad h = 0, \quad p_\mu h^\mu_\nu = 0, \quad b_\mu h^\mu_\nu = 0.
\end{equation}

Despite the violation of the diffeomorphism symmetry, twelve constraints in Eq. (\ref{constrains}) enables only two propagating degrees of freedom, as for the usual Lorentz--invariant graviton. In other words, no graviton mass is produced due to the interaction with the bumblebee field at leading order. The modified dispersion relation (MDR) associated with the physical pole is given by \cite{22}
\begin{equation}
\label{mdr}
    [p^2 + \xi (b \cdot p)^2]h_{\mu\nu} = 0,
\end{equation}
where $b\cdot p = \eta_{\mu\nu}b^{\mu}p^{\mu}$, with $p^{\mu}=(p^{0},{\mathbf p})$ and $b^{\mu}=(b^{0},{\mathbf b})$.
It turns out that the MDR preserves both stability and causality \cite{22}. Also, the respective thermodynamic behavior based on such a modified dispersion relation was addressed in the literature recently by Araújo Filho \cite{aa2021lorentz}.

Also, it is worth mentioning that there exist a correlation of our model under consideration with the Einstein--aether theory. In general lines, the action of this latter case can be written as \cite{eling2006einstein}
\begin{equation}
\label{acttion2}
 S  =  \frac{2}{\kappa^{2}} \int \mathrm{d}x \sqrt{-g} \left\{ -R - K^{\mu\nu}_{\alpha\beta} \nabla_{\mu} u^{\alpha} \nabla_{\nu}u^{\beta}  - \tilde{\lambda}(g_{\mu\nu} u^{\mu} u^{\nu} -1)  \right\},
\end{equation}
where $K^{\mu\nu}_{\alpha\beta} = c_{1} g^{\mu\nu}g_{\alpha\beta} + c_{2} \delta^{\mu}_{\alpha} \delta^{\mu}_{\beta} + c_{3} \delta^{\mu}_{\beta}  \delta^{\nu}_{\alpha} + c_{4} u^{\mu} u^{\nu} g_{\alpha\beta} $.
In the Einstein--aether theory, the action is generally understood to be composed of two primary components: the usual Einstein--Hilbert action, which provides the foundational framework, and a Lagrange multiplier denoted as $\tilde{\lambda}$. Furthermore, covariant terms involving the vector $u^{\mu}$ are incorporated as well, ensuring that they contain at most two derivatives \cite{eling2006einstein}. Based on the kinetic terms, we can argue that Eq. (\ref{acttion}) turns out to be a particular case of Eq. (\ref{acttion2}).


\section{Polarization states}

Studying the polarization of gravitational waves is crucial to gaining insights into the properties of the source and the nature of spacetime. It provides unique information about the motion of massive objects and can reveal any effects encountered during their journey, making it a critical area of research in modern astrophysics \cite{le2017theory,liang2017polarizations,zhang2018velocity,hou2018polarizations,mewes2019signals,liang2022polarizations}. In this sense, we obtain the modified polarization tensor to the free perturbation, satisfying the constraints \eqref{constrains} and the dispersion relation \eqref{mdr}. The source--free equation has the form
\begin{equation}
\label{sourcefreeeom}
    [\Box + \xi (b \cdot \partial)^2]h_{\mu\nu}(x) =0.
\end{equation}
Consider a plane--wave $h_{\mu\nu}=\varepsilon_{\mu\nu}e^{i(\omega t - k z)}$ propagating along the $z$ axis with momentum $p^\mu = (p^0, 0, 0, p^3)$. The constraint $b_\mu h^\mu_\nu = 0$ forbids any state in the direction of the background vector $b^\mu$.

For a timelike $b^\mu = (b^0, 0, 0, 0)$, the polarization tensor reads \begin{equation}
\label{lipolarization}
\varepsilon_{\mu\nu} =
\left(
\begin{array}{cccc}
0 & 0 & 0 & 0 \\
0 & \varepsilon_{11} & \varepsilon_{12} & 0  \\
0 & \varepsilon_{12} & - \varepsilon_{11} & 0 \\
0 & 0 & 0 & 0
\end{array}
\right),
\end{equation}
preserving two Lorentz invariant transversal modes. Nevertheless, the group velocity is modified by
\begin{equation}
    p^0 = \frac{p_3}{\sqrt{1  + (b^0)^2\xi}}.
\end{equation}
Now let us consider a spacelike vector with a spatial component along $b^{\mu}=(0,0,0,b^3)$. The polarization tensor still has the form of Eq. \eqref{lipolarization}, albeit the wave propagates with dispersion relation
\begin{equation}
    p^0 = p^3 \sqrt{1- \xi (b^{3})^{2}}.
\end{equation}
Similar modified dispersion relations were obtained for gauge--invariant containing higher mass dimension operators \citep{26,27}. The group velocity $v_g =\frac{\partial p^0}{\partial p^3}=\sqrt{1- \xi (b^{3})^{2}}<1$.
Therefore, for a timelike or a spacelike $b^\mu$, i.e., with $\vec{b}$ parallel to $\vec{p}$, the transversal characteristic of the gravitational waves is preserved and the wave velocity is slowed by the Lorentz violation.

On the other hand, considering $\vec{b}$ perpendicular to $\vec{p}$, e.g., $b^\mu = (0, 0, b^2, 0)$, we obtain
\begin{equation}
\label{lvpolarization}
\varepsilon_{\mu\nu} =
\left[
\begin{array}{cccc}
-\frac{1}{2}\varepsilon_{11} & -\varepsilon_{13} & 0 &\frac{1}{2} \varepsilon_{11} \\
-\varepsilon_{13} & \varepsilon_{11} & 0 &  \varepsilon_{13}  \\
0 & 0 & 0 & 0 \\
\frac{1}{2}\varepsilon_{11} & \varepsilon_{13} & 0 & -\frac{1}{2}\varepsilon_{11}
\end{array}
\right].
\end{equation}
It is worth mentioning that the polarization tensor \eqref{lvpolarization} has two independent states, $\varepsilon_{11}$ and $\varepsilon_{13}$. The plus state $\varepsilon_{11}$ corresponds to the transversal polarization in the $x$ direction and it is present in the Lorentz invariant wave. Nonetheless, the presence of the cross--state $\varepsilon_{13}$ leads to a longitudinal polarization along the $z$ direction -- a feature forbidden in the Lorentz invariant theory. In other words, the background vector transforms one transversal into a longitudinal degree of freedom. The dispersion relation, in its turn, is kept invariant   
\begin{equation}
    p^0 = p^3.
\end{equation}
Moreover, the line element is modified by
\begin{eqnarray}
    \label{linhao}
    \mathrm{d}s^2 &=& \left(1 - \frac{1}{2}\varepsilon_{11}\right) \mathrm{d}t ^2 - (1-\varepsilon_{11}) \mathrm{d}x^2 - \mathrm{d}y^2 - \left(1 +\frac{1}{2} \varepsilon_{11}\right) \mathrm{d}z^2 + 2\varepsilon_{13} \mathrm{d}x \otimes \mathrm{d}z\nonumber\\
      &-& 2\varepsilon_{13} \mathrm{d}t\otimes \mathrm{d}x +\varepsilon_{11} \mathrm{d}t\otimes \mathrm{d}z .
\end{eqnarray}
In this manner, the Lorentz--violating gravitational waves produce 
contractions and dilatations in the $xz$ plane, and also time deformations.

The effects of these modified gravitational waves on particles can be observed by considering the geodesic deviation equation, which for slowly test particle $U^{\mu}\approx (1,0,0,0)$ reads
\begin{equation}
\label{geodesicdeviation}
\frac{D^2 S^{\mu}}{\mathrm{d}\tau^2}=R^{\mu}_{00\sigma}S^{\sigma},
\end{equation}
where the $S^\mu$ is the displacement vector. From Eq.\eqref{lvpolarization}, the temporal and $y$ components vanish $\frac{D^2 S^{0}}{\mathrm{d}\tau^2}=\frac{D^2 S^{2}}{\mathrm{d}\tau^2}=0$. Although $h_{00}$ is not zero, no time tilde forces arise in this context. In the $x$ direction, Eq.\eqref{geodesicdeviation} gives
\begin{equation}
\label{tidalforcex}
\frac{D^2 S^{1}}{\mathrm{d}\tau^2}=\frac{\omega^2}{2}\Big\{\left(\frac{1}{2}S^0 -S^1 \right)h^{1}_{1} +\left(S^0 -2 S^1 \right)h^{1}_{3}\Big\},
\end{equation}
whereas the tidal force in the longitudinal direction has the form
\begin{equation}
\label{tidalforcez}
\frac{D^2 S^{3}}{\mathrm{d}\tau^2}=-\frac{\omega^2}{2}h^{3}_{1}S^1.
\end{equation}
For two test particles close enough, their events can be considered simultaneous and the corresponding timelike component $S^0$ can be neglected. Note that both plus and cross polarizations contribute to the geodesic deviation in the $x$ direction, while the cross state produces variation along the longitudinal direction only. 
\section{Gravitational waves production in the presence of the bumblebee field}

For the analysis of the production of GW, we add a current $J_{\mu\nu}$ to the homogeneous equation to the graviton field in Eq. \eqref{sourcefreeeom} resulting in
\begin{equation}
    [\Box + \xi (b \cdot \partial)^2]h_{\mu\nu}(x) = J_{\mu\nu}(x).
\end{equation}

The perturbation is determined by 
\begin{equation}
    h_{\mu\nu}(x) = \int \mathrm{d}^4y \; G(x-y) J_{\mu\nu}(y).
\end{equation}

For the modified graviton equation, we can represent the Green function in the momentum space
\begin{equation}
    \Tilde{G}(p) = \frac{1}{p^2 + \xi (b \cdot p)^2},
\end{equation}so that to get $G$ in the configuration space, we may use the inverse Fourier transform. The inversion is made for two particular cases; the first is considering that $b^\mu$ has a timelike configuration $b^\mu = \left(b^0, 0\right)$, in this case, the Green function resumes to
\begin{equation}
    \Tilde{G}(p) = \frac{1}{[1+\xi b_{0}^{2}]p_0^2 - \left|{\bf p}\right|},
\end{equation}and the inversion results to
\begin{equation}
    G(x-y) = \frac{1}{4\pi \sqrt{1+\xi b_{0}^{2}} r}\delta\left[r - \frac{\tau}{\sqrt{1+\xi b_{0}^{2}}}\right].
\end{equation}

The second case is a spacelike configuration $b^\mu = (0,{\bf b})$, which will reduce $\Tilde{G}(p)$ to
\begin{equation}\label{eq:ftGbvec}
    \Tilde{G}(p) = \frac{1}{p_0^2 - (1-\xi |{\bf b}|^2 \cos^2{\Psi})|{\bf p}|},
\end{equation}where $\Psi$ is the angle between ${\bf p}$ and ${\bf b}$, therefore $\cos^2{\Psi} = \cos(\theta) \cos(\theta_b) + \sin(\theta) \sin(\theta_b) \cos(\phi_b - \phi)$, with ($\theta$, $\phi$) and ($\theta_b$, $\phi_b$) the angular coordinates of the trivectors ${\bf p}$ and ${\bf b}$, respectively.

The inverse Fourier transform of Eq. (\ref{eq:ftGbvec}) expanded to the second order of $|{\bf b}|$ is
\begin{equation}
    G(x-y) = G_R(x-y) - G_2(x-y),
\end{equation}{where $G_R(x-y)$ is the retarded Green function and with the second term being given by
\begin{equation}
    \begin{split}
        G_2(x-y) = & \frac{\xi}{8\pi r^3} \Theta(\tau) \times \\
        & \times \left\{(b \cdot r)^2 \left[ \left(\frac{\tau}{r} -1 \right)\delta(\tau - r) + \tau \delta'(\tau - r) \right] + \tau r b^2 \cos({2\theta_b}) \; \delta(\tau - r)\right\}.
    \end{split}
\end{equation}

For the usual Einstein--Hilbert linearized theory, we have the formula for the perturbation
\begin{equation}
    h_{ij}(t,{\bf r}) = \frac{2G}{r}\frac{d^2 I_{ij}}{\mathrm{d}t^2}(t_r),
\end{equation}where $t_r = t - r$ is the retarded time and $I_{ij}$ is the quadrupole momentum defined as
\begin{equation}
    I_{ij}(t) = \int y^i y^j T^{00} \mathrm{d}^3y.
\end{equation}
With $b^\mu = (b^0, 0)$ and considering that $J_{\mu\nu} = 16\pi G T_{\mu\nu}$, the perturbations are written as
\begin{equation}
    h_{ij}(t,{\bf r}) = \frac{2G}{r}\frac{\mathrm{d}^2 I_{ij}}{\mathrm{d}t^2}(t'_r),
\end{equation} 
where the retarded time is modified to $t'_r = t - r\sqrt{1+\xi b_{0}^{2}}$, evidencing that the waves propagates slower. When $b^\mu = (0, {\bf b})$ the perturbation is
\begin{equation}\label{hb}
     h_{ij}(t,{\bf r}) = \frac{2G}{r} \left[ \left( 1-\frac{\xi b^2}{2}\cos{2\theta_b}\right)\frac{\mathrm{d}^2I_{ij}}{\mathrm{d}t^2}(t_r) + \frac{\xi ({\bf b}\cdot {\bf r})^2}{2r} \frac{\mathrm{d}^3 I_{ij}}{\mathrm{d}t^3}(t_r)\right].
\end{equation}
Here, we can see the anisotropy in above solution. Such a feature is well--known within the context of theories modified by the bumblebee vector. The vector ${\bf b}$ selects a preferential direction for the wave propagation. Another peculiarity is that a term involving third derivative of the $I_{ij}$ gives rise to. In addition, terms of dipole are still prohibited due to the momentum conservation.

Now, let us analyze the modifications in the simple binary black hole problem pictured in Fig. \ref{binary}. It represents the movement of a binary system constituted of two masses $m_1$ and $m_2$ in the $xy$ plane. The distances from the center of the reference frame to the respective masses are $r_1$ and $r_2$.

\begin{figure}
    \centering
    \includegraphics[scale=0.4]{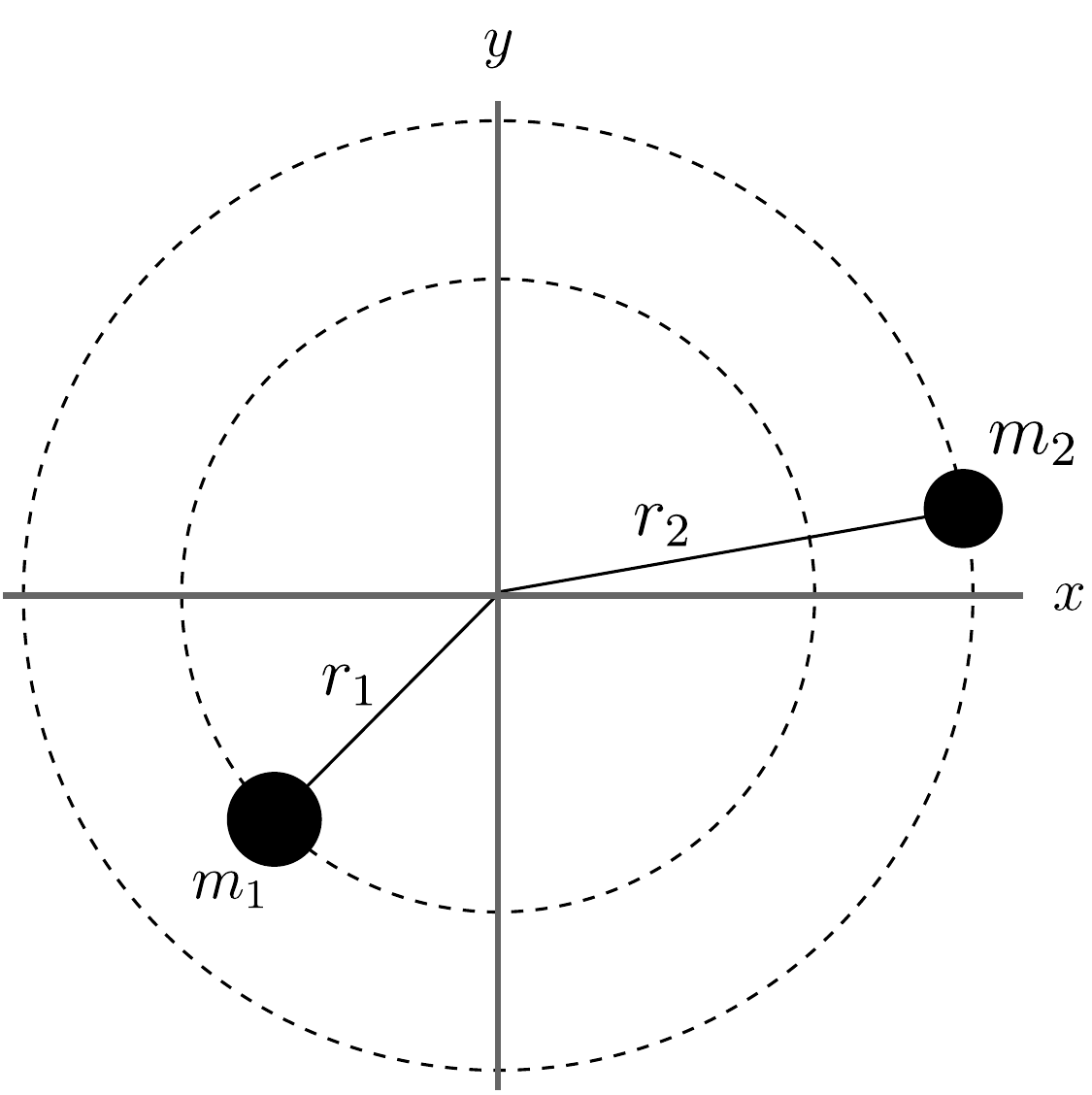}
    \caption{The representation of binary black hole problem.}
    \label{binary}
\end{figure}

Furthermore, the density of energy $T^{00}$ is determined from the formula
\begin{equation}
    T_{00} = \delta(z) [m_1 \delta(x - x_1)\delta(y-y_1) + m_2 \delta(x - x_2)\delta(y-y_2)],
\end{equation}
and the equations of motion of the two particles are
\begin{align}\label{eqmovbin}
    x_1 = \frac{m_2 l_0}{M} \cos{\omega t}, & \qquad y_1 = \frac{m_2 l_0}{M} \sin{\omega t}, \nonumber \\
    x_2 = -\frac{m_1 l_0}{M} \cos{\omega t}, & \qquad y_2 = -\frac{m_1 l_0}{M} \sin{\omega t}.
\end{align}
with $M=m_1+m_2$, $l_0 = r_1+r_2$ and $\omega$ is the angular velocity of the binary. The only non--zero components of $I_{ij}$ are
\begin{align}\label{compI}
    I_{xx} & = \frac{\mu}{2} l_0^2 (1 + \cos{2\omega t}); \nonumber \\
    I_{yy} & = \frac{\mu}{2} l_0^2 (1 - \cos{2\omega t}); \nonumber \\
    I_{xy} & = I_{yx} = \frac{\mu}{2}l_0^2 \sin{2\omega t},
\end{align}
with $\mu = m_1m_2/(m_1+m_2)$. By substituting Eq. \eqref{compI} in Eq. \eqref{hb}, we have
\begin{equation}\label{binmod}
    h_{ij}(t,\mathbf{r}) = 4G\mu l_0^2 \left[\frac{\omega^2}{r} \left(1-\frac{\xi b^2}{2} \cos{(2\theta_b)} \right)A_{ij} + \omega^3\xi (\mathbf{b}\cdot\Hat{\mathbf{r}})^2 B_{ij} \right],
\end{equation}
with the spacial tensors $A_{ij}$ and $B_{ij}$ being defined as
\begin{equation}
    A_{ij} = \left(\begin{array}{ccc}
        -\cos{2\omega t_r} & -\sin{2\omega t_r} & 0 \\
        -\sin{2\omega t_r} & \cos{2\omega t_r} & 0  \\
        0 & 0 & 0
    \end{array}\right), \qquad
    B_{ij} = \left(\begin{array}{ccc}
        \sin{2\omega t_r} & -\cos{2\omega t_r} & 0 \\
        -\cos{2\omega t_r} & -\sin{2\omega t_r} & 0  \\
        0 & 0 & 0
    \end{array} \right).
\end{equation}

From Eq. \eqref{binmod}, we can infer that the frequency of the GW in the presence of the bumblebee field does not change. Besides that, the second term depends of the projection of the bumblebee field in the position vector. Now, we can extract the transverse--traceless part of this solution for the sake of finding the polarization of the waves. Using the projector defined as
\begin{equation}\label{projetorTT}
    \mathcal{P}_{ij,kl} \equiv P_{ik}P_{jl} - \frac{1}{2}P_{ij}P_{kl},
\end{equation}
with $P_{ij} \equiv \delta_{jk} - n_j n_k$ and $n_j = (0,0,1)$ is the unitary vector in the z direction. Therefore, we obtain two polarization states
\begin{align}\label{hTTmod}
    h^{TT}_{xx} &= - 4G\mu l_0^2 \left[\frac{\omega^2}{r} \left(1-\frac{\xi b^2}{2} \cos{(2\theta_b)} \right)\cos{(2\omega t_r)} - \omega^3\xi (\mathbf{b}\cdot\Hat{\mathbf{r}})^2 \sin{(2\omega t_r)} \right], \notag \\
    h^{TT}_{xy} &= - 4G\mu l_0^2 \left[\frac{\omega^2}{r} \left(1-\frac{\xi b^2}{2} \cos{(2\theta_b)} \right)\sin{(2\omega t_r)} + \omega^3\xi (\mathbf{b}\cdot\Hat{\mathbf{r}})^2 \cos{(2\omega t_r)} \right],
\end{align}

which can be rewritten as
\begin{eqnarray}
 h^{TT}_{xx}&=&-A\cos(2\omega t_r -\phi_1)\nonumber\\
 h^{TT}_{xy} &=& -A\sin(2\omega t_r +\phi_1),
\end{eqnarray}
where the resulting amplitude $A$ is given by
\begin{equation}
A=4G\mu l_{0}^2\omega^2 \sqrt{\frac{1}{r^2}\left(1-\frac{\xi b^2}{2}\cos(2\theta_b)\right)^2 +\omega^2 \xi^2 b^4 (\cos \theta_b)^4},
\end{equation}
and the phase difference has the form
\begin{equation}
\phi_1 = \arctan \Big\{\xi\omega r (\vec{b}\cdot \hat{r})\left(1-\frac{\xi b^2}{2}\cos(2\theta_b)\right)^{-1}\Big\}.
\end{equation}

Fig. \ref{hxx_waveform} shows the comparison of $h^{TT}_{xx}$ for three cases, without bumblebee field (Einstein GW), with $\vec{b}$ perpendicular to $\vec{r}$, and with $\vec{b}$ and $\vec{r}$ parallel. It is worth mentioning that the wave frequency is considered low, i.e., before coalescence. Besides that, the parameter $\xi b^2$ was overestimated for the effects to be visible.

\begin{figure}
    \centering
    \includegraphics[scale=0.3]{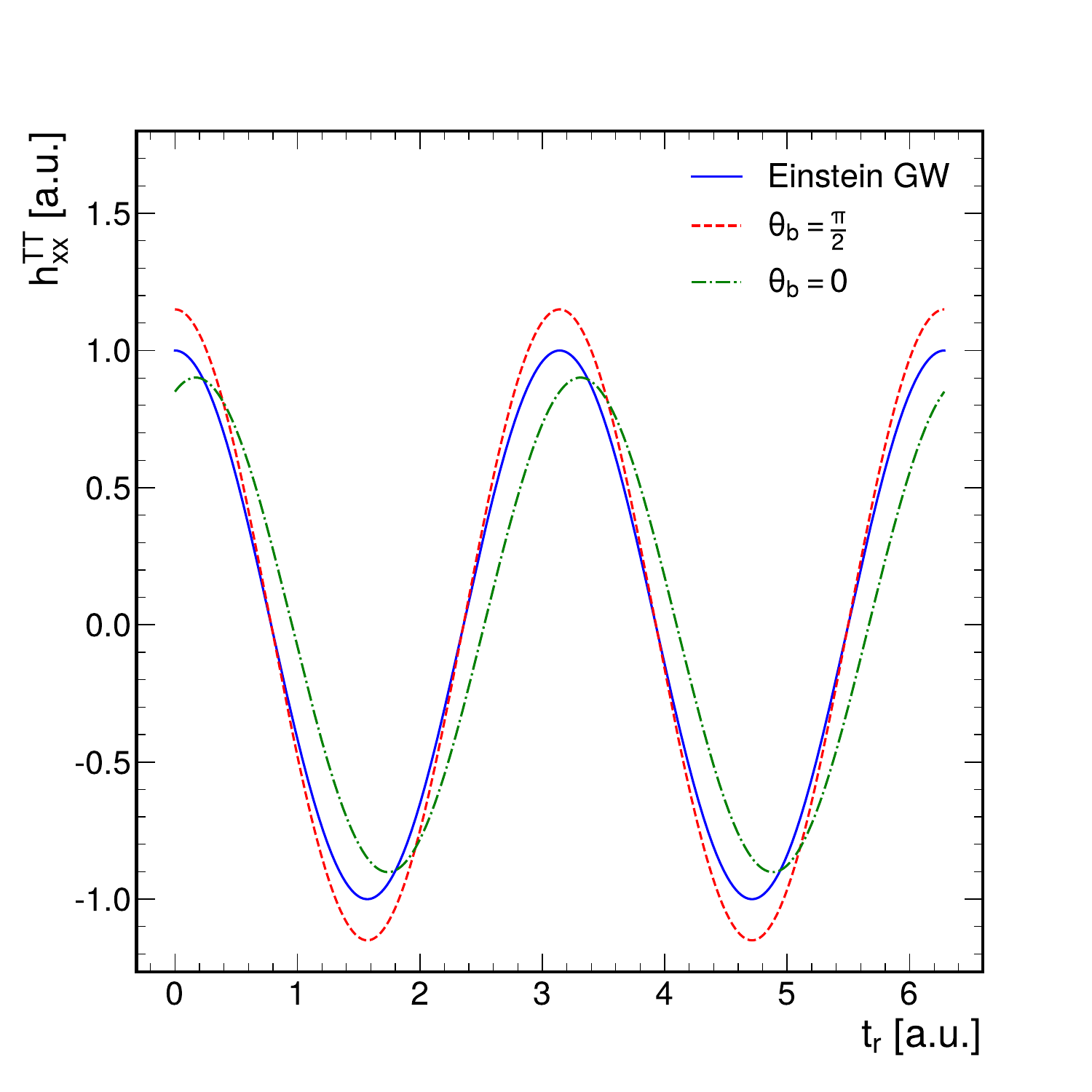}
    \caption{Comparison of $h^{TT}_{xx}$ for three cases, without bumblebee field (Einstein GW -- blue curve), with $\vec{b}$ perpendicular to $\vec{r}$ (red dashed curve), and with $\vec{b}$ and $\vec{r}$ parallel (green dash--doted curve). Note that when $\vec{b}$ and $\vec{r}$ are perpendicular the amplitude is greater and when $\vec{b}$ and $\vec{r}$ are parallel the amplitude is smaller. Besides, a phase difference is observed. In this plot, the frequency is considered low, \textit{i.e.} before coalescence, and the parameter $\xi b^2$ was overestimated for the effects to be visible.}
    \label{hxx_waveform}
\end{figure}

\FloatBarrier

The Lorentz violation modifies the GW amplitude and produces a phase difference which forms an elliptic polarization. This occurs on the $h_{xx}$--$h_{xy}$ plane when $\vec{b}$ and $\vec{r}$ parallel.
Note that both amplitude and phase depend on the relative direction with respect to the background vector.

\FloatBarrier
\begin{figure}
    \centering
    \includegraphics[scale=0.3]{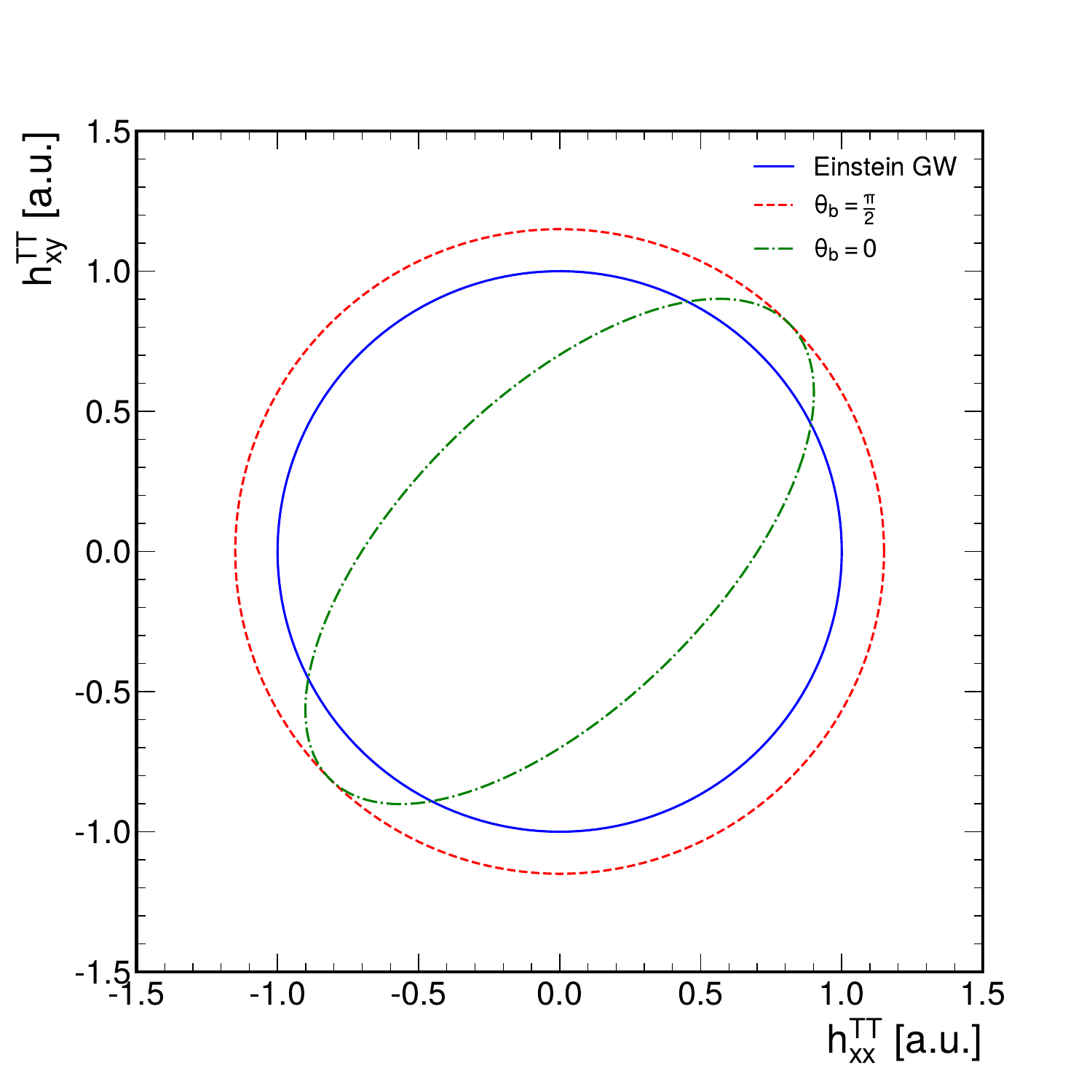}
    \caption{Polarization of the GW in the $h_{xx}$--$h_{xy}$ plane for three cases, without bumblebee field (Einstein GW -- blue curve), with $\vec{b}$ perpendicular to $\vec{r}$ (red dashed curve), and with $\vec{b}$ and $\vec{r}$ parallel (green dash--doted curve). It is observed that when $\vec{b}$ and $\vec{r}$ are parallel the GW is elliptically polarized. When $\vec{b}$ and $\vec{r}$ are perpendicular, the polarization is circular like the Einstein GW, but with greater amplitude.}
    \label{pol_xxvsxy}
\end{figure}

At leading order, the correction in the amplitude due to the Lorentz violation has the form
\begin{equation}
\frac{\delta A}{A_{LI}}\approx -\frac{1}{2}\xi b^2 \cos 2\theta_b.
\end{equation}
Therefore, for $\vec{k}$ parallel $(\theta_b =0)$ and anti--parallel $(\theta_b =\pi)$ to $\vec{b}$, the Lorentz violation reduces the amplitude to its minimal, whereas for $\vec{k}$ perpendicular to $\vec{b}$, the correction attains its maximum value, as shown In Fig. \eqref{hxx_waveform}. Also, in Fig. \ref{pol_xxvsxy}, we display the polarization analysis of gravitational waves in the $h_{xx}$–$h_{xy}$ plane reveals distinct behaviors in three scenarios: without the bumblebee field (Einstein gravitational waves denoted by the blue curve), with the vector field $\vec{b}$ perpendicular to $\vec{r}$ (illustrated by the red dashed curve), and with $\vec{b}$ and $\vec{r}$ oriented in parallel (depicted by the green dash--dotted curve). It is noteworthy that in the case where $\vec{b}$ and $\vec{r}$ align, the gravitational wave exhibits elliptical polarization. Conversely, when $\vec{b}$ and $\vec{r}$ are perpendicular, the polarization mirrors the circular pattern characteristic of Einstein gravitational waves, albeit with a heightened amplitude.


\section{Estimation of bounds}

In order to set upper bounds for the Lorentz violation parameter, we can compare the group velocity that comes from \eqref{sourcefreeeom}\begin{equation}
    \label{grupovelo}
    u_g = \left[\frac{\xi b^0 |\Vec{b}| \cos\Psi \pm \sqrt{1 + \xi (b^0)^2 - \xi |\Vec{b}|^2 \cos^2\Psi}}{1+\xi (b^0)^2}\right],
\end{equation} 
\noindent with the one obtained in \cite{will1998bounding}, which is the same used by the LIGO collaboration \cite{abbott1}. Therefore, using the group velocity for the massive graviton\begin{equation}
    v_g \thicksim  1 - \frac{1}{2}\left(\frac{m_g}{k_g}\right)^2,
\end{equation}

\noindent where $m_g$ is the upper bound for the assumed mass of the graviton and $k_g$, is the graviton energy. The LIGO and Virgo collaborations reported that the peak of the gravitational wave signal  detected in event GW150914 has a frequency of $\nu$ = 150 Hz \cite{abbott1} so that the energy is $k_g = h\nu \thicksim 6.024 \hspace{0.02cm} \cdot \hspace{0.02cm} 10^{-13}$ eV (for $h = 4.136 \hspace{0.02cm} \cdot \hspace{0.02cm} 10^{-15}$ eV). Besides that, the lower bound for the graviton mass found by them is given by $m_g < 1.20 \hspace{0.02cm} \cdot \hspace{0.02cm} 10^{-22}$ eV so that\begin{equation}
    \Delta v_g = c - v_g < 2,0 \hspace{0.02cm} \cdot \hspace{0.02cm} 10^{-20},
\end{equation}  

\noindent which is the value to be used to estimate the bounds of the Lorentz violating parameters. This, we can perform the same estimate for the group velocity \eqref{grupovelo} found in this work. First, let $b^\mu$ be space-like\begin{equation}
    u_g \thicksim 1 - \frac{1}{2} \xi |\vec{b}|^2 \cos^2{\Psi},
\end{equation}

\noindent so that\begin{equation}
    \begin{split}
        \Delta u_g & = c - u_g \\ & = \frac{1}{2} \xi |\vec{b}|^2 \cos^2{\Psi}.
    \end{split}
\end{equation}

\noindent For the case where $\vec{b}$ e $\vec{p}$ are orthogonal,\begin{equation}
    \xi|\vec{b}|^2 = 2\Delta u_g < 4 . 10^{-20}.
\end{equation}

\noindent For $b^\mu$ time--like, we have\begin{equation}
    \begin{split}
        \xi(b_0)^2 & \thicksim \frac{2\Delta v_g}{1 -2 \Delta v_g} \\ & \thicksim 2 \Delta v_g,
    \end{split}
\end{equation}

\noindent which is the same value found for the space--like parameter.

Within the MPE, many physicists are developing theories and experiments where the basic premise is that tiny Lorentz and CPT violations can be observed in nature. In this way, Kostelecký and its collaborators made a compilation of possible experimental results so the scientific community could investigate \cite{kostelecky2011data}. Some examples of measuring Lorentz and CPT violation coefficients are: neutrino oscillations \cite{aharmim2018tests, barabash2018final, adey2018search}, kaons oscillations \cite{babusci2014test, di2010cpt}, production and decay of top quark \cite{abazov2012vm}, electroweak sector \cite{sytema2016sytema, muller2013se}, clock--comparison experiments \cite{pruttivarasin2015michelson, botermann2014test, hohensee2013limits}, photon sector \cite{kislat2018constraints, parker2015bounds}, gravitational sector \cite{shao2018combined, abbott2017gw170817, shao2018limits, casana2018exact, bourgoin2017lorentz}. 

Therefore, the modifications in the polarization states determined here can be found by gravitational wave detection experiments. Between LIGO and LISA, the one that has a greater perspective of finding such a modification is LISA, as it will be orbiting the planet Earth so its orientation in relation to a theoretical background vector will have more variations than LIGO. This same idea is used in clock comparison experiments, where better results are found when clocks are orbiting planet Earth (\textit{e.g.} satellites or international space stations). However, as LISA will only be launched in 2034, we can think of some modifications in the experiments that already exist here on Earth, so that the pendulums that form the test masses can be adjusted so that the polarization modified due to a theoretical vector background can be measured.

Based on the LIGO measurements \cite{abbott1}, we can estimate the bound for the amplitude. To do so, we suppose the contribution of the Lorentz violation $\delta A$ is less than the observational error. Such a procedure allows us to estimate an upper bound at the level of $\left|\xi b^2\right|\leq 2.55\times10^{-19}.$

\section{Conclusion}

In this work, we investigated the effects of the Lorentz symmetry breaking due to the presence of a non--zero VEV for a vector field. We used the simplest model in literature, the so--called bumblebee model. We expanded the density Lagrangian of the theory up to the second order of $h_{\mu\nu}$, and we observed the modified equation of motion to the graviton. Hence, we obtained the modified dispersion relation and the corresponding plane wave solution. We also showed that the free graviton had only two degrees of freedom, and we explicitly determined the new polarization tensor dependent on the background field. Furthermore, despite the coupling with the bumblebee field, no massive mode was found for the graviton.

Therefore, the modified wave equation was solved in terms of the Fourier modes. Taking into account the constrains $p_\mu h^{\mu\nu}=0$ and $b_\mu h^{\mu\nu}=0$, we got modifications in polarization states. For the case where $b^\mu$ was timelike or $\vec{b}$ was in the same direction of the wave, no modifications appeared in the polarization tensor. Nevertheless, there existed modifications in the group velocity. In addition, for the case where $b^\mu$ was spacelike and $\vec{b}$ was orthogonal to the momentum, there existed significant modifications, such that one polarization mode was in a orthogonal plane with respect to $\vec{b}$ and the another one was longitudinal with respect to the wave momentum. Besides that, we obtained the following upper bound for the bumblebee parameter, $\xi|\vec{b}|^2 < 4 . 10^{-20}$. This was done by comparing the group velocity obtained here with the group velocity for the massive graviton and using experimental values of the GW150914 event.

Thereby, we sought for the effects of the spontaneous Lorentz breaking in the production of the gravitational waves. Assuming the presence of a matter source, we calculated the Green function and the corresponding solution for the graviton field in two special configurations. In the first one, we considered a timelike configuration for $b^\mu$. In this case, only the velocity of propagation of the wave was smaller by a factor of $\sqrt{1+\xi b_{0}^{2}}$. In the second case, we regarded a spacelike configuration for the bumblebee VEV. The corrections to the quadrupole formula showed the existence of anisotropy in the solution, showing an explicit dependency in the relative direction between the background field and the position vector. The frequency of the wave did not change, once the tensor $I_{ij}$ remained the same of the usual Einstein--Hilbert theory. In the circular orbit binary solution, we showed that the polarizations were not circular in contrast with the usual solution. Finally, the amplitude changed by a factor of $1-\xi b^2/2$.

\section{Acknowledgments}

\hspace{0.5cm}The authors would like to thank the Carlos Chagas Filho Foundation for Research Support in the State of Rio de Janeiro (FAPERJ), the Coordination of Superior Level Staff Improvement (CAPES), and the National Council for Scientific and Technological Development (CNPq) for financial support. Particularly, A. A. Araújo Filho is supported by Conselho Nacional de Desenvolvimento Cient\'{\i}fico e Tecnol\'{o}gico (CNPq) and Fundação de Apoio à Pesquisa do Estado da Paraíba (FAPESQ), project No. 150891/2023-7.

\section{Data Availability Statement}

Data Availability Statement: No Data associated in the manuscript


\bibliographystyle{ieeetr}
\bibliography{main}

\begin{thebibliography}{10}

\bibitem{1}
V.~A. Kosteleck{\`y} and S.~Samuel, ``Spontaneous breaking of lorentz symmetry
  in string theory,'' {\em Physical Review D}, vol.~39, no.~2, p.~683, 1989.

\bibitem{1.2}
V.~A. Kosteleck{\`y} and S.~Samuel, ``Phenomenological gravitational
  constraints on strings and higher-dimensional theories,'' {\em Physical
  Review Letters}, vol.~63, no.~3, p.~224, 1989.

\bibitem{2}
J.~Alfaro, H.~A. Morales-Tecotl, and L.~F. Urrutia, ``Quantum gravity
  corrections to neutrino propagation,'' {\em Physical Review Letters},
  vol.~84, no.~11, p.~2318, 2000.

\bibitem{2.1}
J.~Alfaro, H.~A. Morales-Tecotl, and L.~F. Urrutia, ``Loop quantum gravity and
  light propagation,'' {\em Physical Review D}, vol.~65, no.~10, p.~103509,
  2002.

\bibitem{3}
S.~M. Carroll, J.~A. Harvey, V.~A. Kosteleck{\`y}, C.~D. Lane, and T.~Okamoto,
  ``Noncommutative field theory and lorentz violation,'' {\em Physical Review
  Letters}, vol.~87, no.~14, p.~141601, 2001.

\bibitem{araujo2023thermodynamics}
A.~Ara{\'u}jo~Filho, S.~Zare, P.~Porf{\'\i}rio, J.~K{\v{r}}{\'\i}{\v{z}}, and
  H.~Hassanabadi, ``Thermodynamics and evaporation of a modified schwarzschild
  black hole in a non--commutative gauge theory,'' {\em Physics Letters B},
  vol.~838, p.~137744, 2023.

\bibitem{heidari2023gravitational}
N.~Heidari, H.~Hassanabadi, A.~A. Ara{\'u}jo~Filho, J.~Kriz, S.~Zare, and
  P.~Porf{\'\i}rio, ``Gravitational signatures of a non--commutative stable
  black hole,'' {\em Physics of the Dark Universe}, p.~101382, 2023.

\bibitem{4}
P.~Ho{\v{r}}ava, ``Quantum gravity at a lifshitz point,'' {\em Physical Review
  D}, vol.~79, no.~8, p.~084008, 2009.

\bibitem{5}
A.~G. Cohen and S.~L. Glashow, ``Very special relativity,'' {\em Physical
  review letters}, vol.~97, no.~2, p.~021601, 2006.

\bibitem{furtado2023thermal}
A.~A. Ara{\'u}jo~Filho, J.~Furtado, H.~Hassanabadi, and J.~Reis, ``Thermal
  analysis of photon-like particles in rainbow gravity,'' {\em Physics of the
  Dark Universe}, vol.~42, p.~101310, 2023.

\bibitem{6}
D.~Colladay and V.~A. Kosteleck{\`y}, ``Cpt violation and the standard model,''
  {\em Physical Review D}, vol.~55, no.~11, p.~6760, 1997.

\bibitem{6.1}
D.~Colladay and V.~A. Kosteleck{\`y}, ``Cpt violation and the standard model,''
  {\em Physical Review D}, vol.~55, no.~11, p.~6760, 1997.

\bibitem{7}
R.~Jackiw and V.~A. Kosteleck{\`y}, ``Radiatively induced lorentz and cpt
  violation in electrodynamics,'' {\em Physical Review Letters}, vol.~82,
  no.~18, p.~3572, 1999.

\bibitem{araujo2021thermodynamic}
A.~A. Ara{\'u}jo~Filho and R.~Maluf, ``Thermodynamic properties in
  higher-derivative electrodynamics,'' {\em Brazilian Journal of Physics},
  vol.~51, pp.~820--830, 2021.

\bibitem{araujo2021higher}
A.~A. Ara{\'u}jo~Filho and A.~Y. Petrov, ``Higher-derivative lorentz-breaking
  dispersion relations: a thermal description,'' {\em The European Physical
  Journal C}, vol.~81, no.~9, p.~843, 2021.

\bibitem{araujo2022particles}
A.~A. Ara{\'u}jo~Filho, ``Particles in loop quantum gravity formalism: a
  thermodynamical description,'' {\em Annalen der Physik}, p.~2200383, 2022.

\bibitem{araujo2021bouncing}
A.~A. Ara{\'u}jo~Filho and A.~Y. Petrov, ``Bouncing universe in a heat bath,''
  {\em International Journal of Modern Physics A}, vol.~36, no.~34n35,
  p.~2150242, 2021.

\bibitem{araujo2022thermal}
A.~A. Ara{\'u}jo~Filho, {\em Thermal aspects of field theories}.
\newblock Amazon. com, 2022.

\bibitem{8}
V.~A. Kosteleck{\`y} and M.~Mewes, ``Lorentz and cpt violation in neutrinos,''
  {\em Physical Review D}, vol.~69, no.~1, p.~016005, 2004.

\bibitem{9}
R.~Bluhm, V.~A. Kosteleck{\`y}, and C.~D. Lane, ``Cpt and lorentz tests with
  muons,'' {\em Physical Review Letters}, vol.~84, no.~6, p.~1098, 2000.

\bibitem{ww1}
A.~Ara{\'u}jo~Filho, H.~Hassanabadi, J.~Reis, and L.~Lisboa-Santos,
  ``Thermodynamics of a quantum ring modified by lorentz violation,'' {\em
  Physica Scripta}, vol.~98, no.~6, p.~065943, 2023.

\bibitem{10}
R.~Bluhm, V.~A. Kosteleck{\`y}, and N.~Russell, ``Cpt and lorentz tests in
  hydrogen and antihydrogen,'' {\em Physical Review Letters}, vol.~82, no.~11,
  p.~2254, 1999.

\bibitem{filho2022thermodynamics}
A.~A. Ara{\'u}jo~Filho, ``Thermodynamics of massless particles in curved
  spacetime,'' {\em arXiv preprint arXiv:2201.00066}, 2022.

\bibitem{li2017application}
G.-P. Li, J.~Pu, Q.-Q. Jiang, and X.-T. Zu, ``An application of
  lorentz-invariance violation in black hole thermodynamics,'' {\em The
  European Physical Journal C}, vol.~77, pp.~1--10, 2017.

\bibitem{cambiaso2012massive}
M.~Cambiaso, R.~Lehnert, and R.~Potting, ``Massive photons and lorentz
  violation,'' {\em Physical Review D}, vol.~85, no.~8, p.~085023, 2012.

\bibitem{ww2}
A.~A. Ara{\'u}jo~Filho, J.~Reis, and S.~Ghosh, ``Quantum gases on a torus,''
  {\em International Journal of Geometric Methods in Modern Physics},
  p.~2350178, 2023.

\bibitem{ww3}
A.~Ara{\'u}jo~Filho and J.~Reis, ``How does geometry affect quantum gases?,''
  {\em International Journal of Modern Physics A}, vol.~37, no.~11n12,
  p.~2250071, 2022.

\bibitem{ww4}
A.~A. Ara{\'u}jo~Filho, J.~A. A.~S. Reis, and S.~Ghosh, ``Fermions on a torus
  knot,'' {\em The European Physical Journal Plus}, vol.~137, no.~5, p.~614,
  2022.

\bibitem{11}
V.~A. Kosteleck{\`y} and N.~Russell, ``Data tables for lorentz and c p t
  violation,'' {\em Reviews of Modern Physics}, vol.~83, no.~1, p.~11, 2011.

\bibitem{bluhm2008spontaneous}
R.~Bluhm, S.-H. Fung, and V.~A. Kosteleck{\`y}, ``Spontaneous lorentz and
  diffeomorphism violation, massive modes, and gravity,'' {\em Physical Review
  D}, vol.~77, no.~6, p.~065020, 2008.

\bibitem{12}
B.~P. Abbott, R.~Abbott, T.~Abbott, M.~Abernathy, F.~Acernese, K.~Ackley,
  C.~Adams, T.~Adams, P.~Addesso, R.~Adhikari, {\em et~al.}, ``Observation of
  gravitational waves from a binary black hole merger,'' {\em Physical review
  letters}, vol.~116, no.~6, p.~061102, 2016.

\bibitem{13}
B.~P. Abbott, R.~Abbott, T.~Abbott, M.~Abernathy, F.~Acernese, K.~Ackley,
  C.~Adams, T.~Adams, P.~Addesso, R.~Adhikari, {\em et~al.}, ``Gw150914: The
  advanced ligo detectors in the era of first discoveries,'' {\em Physical
  review letters}, vol.~116, no.~13, p.~131103, 2016.

\bibitem{aa2023analysis}
A.~A. Ara{\'u}jo~Filho, ``Analysis of a regular black hole in verlinde’s
  gravity,'' {\em Classical and Quantum Gravity}, vol.~41, no.~1, p.~015003,
  2023.

\bibitem{filho2024implications}
A.~A. Ara{\'u}jo~Filho, ``Implications of a simpson--visser solution in
  verlinde’s framework,'' {\em The European Physical Journal C}, vol.~84,
  no.~1, p.~73, 2024.

\bibitem{15}
R.~Bluhm and V.~A. Kosteleck{\`y}, ``Spontaneous lorentz violation,
  nambu-goldstone modes, and gravity,'' {\em Physical Review D}, vol.~71,
  no.~6, p.~065008, 2005.

\bibitem{16}
R.~Bluhm, S.-H. Fung, and V.~A. Kosteleck{\`y}, ``Spontaneous lorentz and
  diffeomorphism violation, massive modes, and gravity,'' {\em Physical Review
  D}, vol.~77, no.~6, p.~065020, 2008.

\bibitem{maluf2019antisymmetric}
R.~Maluf, A.~Ara{\'u}jo~Filho, W.~Cruz, and C.~Almeida, ``Antisymmetric tensor
  propagator with spontaneous lorentz violation,'' {\em Europhysics Letters},
  vol.~124, no.~6, p.~61001, 2019.

\bibitem{sagi2010propagation}
E.~Sagi, ``Propagation of gravitational waves in the generalized
  tensor-vector-scalar theory,'' {\em Physical Review D}, vol.~81, no.~6,
  p.~064031, 2010.

\bibitem{liang2022polarizations}
D.~Liang, R.~Xu, X.~Lu, and L.~Shao, ``Polarizations of gravitational waves in
  the bumblebee gravity model,'' {\em Physical Review D}, vol.~106, no.~12,
  p.~124019, 2022.

\bibitem{17}
V.~A. Kosteleck{\`y}, ``Gravity, lorentz violation, and the standard model,''
  {\em Physical Review D}, vol.~69, no.~10, p.~105009, 2004.

\bibitem{nascimento2022vacuum}
A.~A. Ara{\'u}jo~Filho, J.~R. Nascimento, A.~Y. Petrov, and P.~J.
  Porf{\'\i}rio, ``Vacuum solution within a metric-affine bumblebee gravity,''
  {\em Physical Review D}, vol.~108, no.~8, p.~085010, 2023.

\bibitem{18}
R.~Maluf, V.~Santos, W.~Cruz, and C.~Almeida, ``Matter-gravity scattering in
  the presence of spontaneous lorentz violation,'' {\em Physical Review D},
  vol.~88, no.~2, p.~025005, 2013.

\bibitem{hassanabadi2023gravitational}
A.~A. Ara{\'u}jo~Filho, H.~Hassanabadi, N.~Heidari, J.~Kriz, and S.~Zare,
  ``Gravitational traces of bumblebee gravity in metric--affine formalism,''
  {\em Classical and Quantum Gravity}, vol.~41, no.~5, p.~055003, 2024.

\bibitem{19}
A.~Mart{\'\i}n-Ruiz and C.~Escobar, ``Local effects of the quantum vacuum in
  lorentz-violating electrodynamics,'' {\em Physical Review D}, vol.~95, no.~3,
  p.~036011, 2017.

\bibitem{20}
C.~Hernaski, ``Quantization and stability of bumblebee electrodynamics,'' {\em
  Physical Review D}, vol.~90, no.~12, p.~124036, 2014.

\bibitem{22}
R.~Maluf, C.~Almeida, R.~Casana, and M.~Ferreira~Jr, ``Einstein-hilbert
  graviton modes modified by the lorentz-violating bumblebee field,'' {\em
  Physical Review D}, vol.~90, no.~2, p.~025007, 2014.

\bibitem{aa2021lorentz}
A.~A. Ara{\'u}jo~Filho, ``Lorentz-violating scenarios in a thermal reservoir,''
  {\em The European Physical Journal Plus}, vol.~136, no.~4, pp.~1--14, 2021.

\bibitem{eling2006einstein}
C.~Eling, T.~Jacobson, and D.~Mattingly, ``Einstein-aether theory,'' in {\em
  Deserfest}, pp.~163--179, World Scientific, 2006.

\bibitem{le2017theory}
A.~Le~Tiec and J.~Novak, ``Theory of gravitational waves,'' in {\em An Overview
  of Gravitational Waves: Theory, Sources and Detection}, pp.~1--41, World
  Scientific, 2017.

\bibitem{liang2017polarizations}
D.~Liang, Y.~Gong, S.~Hou, Y.~Liu, {\em et~al.}, ``Polarizations of
  gravitational waves in f (r) gravity,'' {\em Physical Review D}, vol.~95,
  no.~10, p.~104034, 2017.

\bibitem{zhang2018velocity}
P.-M. Zhang, C.~Duval, G.~Gibbons, and P.~Horvathy, ``Velocity memory effect
  for polarized gravitational waves,'' {\em Journal of Cosmology and
  Astroparticle Physics}, vol.~2018, no.~05, p.~030, 2018.

\bibitem{hou2018polarizations}
S.~Hou, Y.~Gong, and Y.~Liu, ``Polarizations of gravitational waves in
  horndeski theory,'' {\em The European Physical Journal C}, vol.~78, no.~5,
  p.~378, 2018.

\bibitem{mewes2019signals}
M.~Mewes, ``Signals for lorentz violation in gravitational waves,'' {\em
  Physical Review D}, vol.~99, no.~10, p.~104062, 2019.

\bibitem{26}
V.~A. Kosteleck{\`y} and M.~Mewes, ``Testing local lorentz invariance with
  gravitational waves,'' {\em Physics Letters B}, vol.~757, pp.~510--514, 2016.

\bibitem{27}
V.~A. Kosteleck{\`y} and M.~Mewes, ``Lorentz and diffeomorphism violations in
  linearized gravity,'' {\em Physics Letters B}, vol.~779, pp.~136--142, 2018.

\bibitem{will1998bounding}
C.~M. Will, ``Bounding the mass of the graviton using gravitational-wave
  observations of inspiralling compact binaries,'' {\em Physical Review D},
  vol.~57, no.~4, p.~2061, 1998.

\bibitem{abbott1}
B.~P. Abbott, R.~Abbott, T.~Abbott, M.~Abernathy, F.~Acernese, K.~Ackley,
  C.~Adams, T.~Adams, P.~Addesso, R.~Adhikari, {\em et~al.}, ``Observation of
  gravitational waves from a binary black hole merger,'' {\em Physical review
  letters}, vol.~116, no.~6, p.~061102, 2016.

\bibitem{kostelecky2011data}
V.~A. Kosteleck{\`y} and N.~Russell, ``Data tables for lorentz and c p t
  violation,'' {\em Reviews of Modern Physics}, vol.~83, no.~1, p.~11, 2011.

\bibitem{aharmim2018tests}
B.~Aharmim, S.~Ahmed, A.~Anthony, N.~Barros, E.~Beier, A.~Bellerive,
  B.~Beltran, M.~Bergevin, S.~Biller, E.~Blucher, {\em et~al.}, ``Tests of
  lorentz invariance at the sudbury neutrino observatory,'' {\em Physical
  Review D}, vol.~98, no.~11, p.~112013, 2018.

\bibitem{barabash2018final}
A.~Barabash, P.~Belli, R.~Bernabei, F.~Cappella, V.~Caracciolo, R.~Cerulli,
  D.~Chernyak, F.~Danevich, S.~d’Angelo, A.~Incicchitti, {\em et~al.},
  ``Final results of the aurora experiment to study 2 $\beta$ decay of cd 116
  with enriched cd 116 wo 4 crystal scintillators,'' {\em Physical Review D},
  vol.~98, no.~9, p.~092007, 2018.

\bibitem{adey2018search}
D.~Adey, F.~An, A.~Balantekin, H.~Band, M.~Bishai, S.~Blyth, D.~Cao, G.~Cao,
  J.~Cao, J.~Chang, {\em et~al.}, ``Search for a time-varying electron
  antineutrino signal at daya bay,'' {\em Physical Review D}, vol.~98, no.~9,
  p.~092013, 2018.

\bibitem{babusci2014test}
D.~Babusci, I.~Balwierz-Pytko, G.~Bencivenni, C.~Bloise, F.~Bossi,
  P.~Branchini, A.~Budano, L.~C. Balkest{\aa}hl, G.~Capon, F.~Ceradini, {\em
  et~al.}, ``Test of cpt and lorentz symmetry in entangled neutral kaons with
  the kloe experiment,'' {\em Physics Letters B}, vol.~730, pp.~89--94, 2014.

\bibitem{di2010cpt}
A.~Di~Domenico, K.~collaboration, {\em et~al.}, ``Cpt symmetry and quantum
  mechanics tests in the neutral kaon system at kloe,'' {\em Foundations of
  Physics}, vol.~40, no.~7, pp.~852--866, 2010.

\bibitem{abazov2012vm}
V.~Abazov, ``Vm abazov et al.(d0 collaboration), phys. rev. lett. 108, 151804
  (2012).,'' {\em Phys. Rev. Lett.}, vol.~108, p.~151804, 2012.

\bibitem{sytema2016sytema}
A.~Sytema, ``A. sytema et al., phys. rev. c 94, 025503 (2016).,'' {\em Phys.
  Rev. C}, vol.~94, p.~025503, 2016.

\bibitem{muller2013se}
S.~M{\"u}ller, ``Se m{\"u}ller et al., phys. rev. d 88, 071901 (r)(2013),''
  {\em Phys. Rev. D}, vol.~88, p.~071901, 2013.

\bibitem{pruttivarasin2015michelson}
T.~Pruttivarasin, M.~Ramm, S.~Porsev, I.~Tupitsyn, M.~Safronova, M.~Hohensee,
  and H.~H{\"a}ffner, ``Michelson--morley analogue for electrons using trapped
  ions to test lorentz symmetry,'' {\em Nature}, vol.~517, no.~7536, p.~592,
  2015.

\bibitem{botermann2014test}
B.~Botermann, D.~Bing, C.~Geppert, G.~Gwinner, T.~W. H{\"a}nsch, G.~Huber,
  S.~Karpuk, A.~Krieger, T.~K{\"u}hl, W.~N{\"o}rtersh{\"a}user, {\em et~al.},
  ``Test of time dilation using stored li+ ions as clocks at relativistic
  speed,'' {\em Physical review letters}, vol.~113, no.~12, p.~120405, 2014.

\bibitem{hohensee2013limits}
M.~Hohensee, N.~Leefer, D.~Budker, C.~Harabati, V.~Dzuba, and V.~Flambaum,
  ``Limits on violations of lorentz symmetry and the einstein equivalence
  principle using radio-frequency spectroscopy of atomic dysprosium,'' {\em
  Physical review letters}, vol.~111, no.~5, p.~050401, 2013.

\bibitem{kislat2018constraints}
F.~Kislat, ``Constraints on lorentz invariance violation from optical
  polarimetry of astrophysical objects,'' {\em Symmetry}, vol.~10, no.~11,
  p.~596, 2018.

\bibitem{parker2015bounds}
S.~R. Parker, M.~Mewes, F.~N. Baynes, and M.~E. Tobar, ``Bounds on higher-order
  lorentz-violating photon sector coefficients from an asymmetric optical ring
  resonator experiment,'' {\em Physics Letters A}, vol.~379, no.~42,
  pp.~2681--2684, 2015.

\bibitem{shao2018combined}
C.-G. Shao, Y.-F. Chen, Y.-J. Tan, S.-Q. Yang, J.~Luo, M.~E. Tobar, J.~Long,
  E.~Weisman, and A.~Kostelecky, ``Combined search for a lorentz-violating
  force in short-range gravity varying as the inverse sixth power of
  distance,'' {\em arXiv preprint arXiv:1812.11123}, 2018.

\bibitem{abbott2017gw170817}
B.~P. Abbott, R.~Abbott, T.~Abbott, F.~Acernese, K.~Ackley, C.~Adams, T.~Adams,
  P.~Addesso, R.~Adhikari, V.~Adya, {\em et~al.}, ``Gw170817: observation of
  gravitational waves from a binary neutron star inspiral,'' {\em Physical
  Review Letters}, vol.~119, no.~16, p.~161101, 2017.

\bibitem{shao2018limits}
C.-G. Shao, Y.-F. Chen, R.~Sun, L.-S. Cao, M.-K. Zhou, Z.-K. Hu, C.~Yu, and
  H.~M{\"u}ller, ``Limits on lorentz violation in gravity from worldwide
  superconducting gravimeters,'' {\em Physical Review D}, vol.~97, no.~2,
  p.~024019, 2018.

\bibitem{casana2018exact}
R.~Casana, A.~Cavalcante, F.~Poulis, and E.~Santos, ``Exact schwarzschild-like
  solution in a bumblebee gravity model,'' {\em Physical Review D}, vol.~97,
  no.~10, p.~104001, 2018.

\bibitem{bourgoin2017lorentz}
A.~Bourgoin, C.~Le~Poncin-Lafitte, A.~Hees, S.~Bouquillon, G.~Francou, and
  M.-C. Angonin, ``Lorentz symmetrpassos2017lorentzy violations from
  matter-gravity couplings with lunar laser ranging,'' {\em Physical review
  letters}, vol.~119, no.~20, p.~201102, 2017.

\end{thebibliography}

\end{document}